\documentclass[12pt,letterpaper]{article}
\pdfoutput=1
\usepackage{jheppub}
\usepackage{amsfonts, amsthm}
\usepackage[english]{babel}
\usepackage[utf8]{inputenc}
\usepackage{slashed}
\usepackage{mathrsfs}
\usepackage{amssymb}
\usepackage{color}
\hypersetup{unicode}
\usepackage{natbib}
\usepackage{arydshln}

\newcommand{\eq}{\begin{equation}}
\newcommand{\feq}{\end{equation}}
\newcommand{\eqn}{\begin{eqnarray}}
\newcommand{\feqn}{\end{eqnarray}}

\newcommand{\be}{\begin{equation}}
\newcommand{\ee}{\end{equation}}
\newcommand{\ma}[1]{\mbox{$\mathcal{#1}$}}

\newcommand{\mrm}[1]{\mbox{$\mathrm{#1}$}}

\title{AdS$_3$ from M-branes at conical singularities}

\author[a,b]{Giuseppe Dibitetto}
\author[c]{Nicol\`o Petri}

\affiliation[a]{Dipartimento di Fisica e Astronomia, Universit\`a di Padova, via Marzolo 8, 35131 Padova, Italy}
\affiliation[b]{INFN, Sezione di Padova, via Marzolo 8, 35131 Padova, Italy}
\affiliation[c]{Department of Physics, University of Oviedo, Avda. Federico Garcia Lorca s/n, 33007 Oviedo,
Spain}

\emailAdd{giuseppe.dibitetto@unipd.it}
\emailAdd{petrinicolo@uniovi.es}

\abstract{M-theory is known to possess supersymmetric solutions where the geometry is $\mathrm{AdS}_3\times S^3\times S^3$ warped over a Riemann surface $\Sigma_{2}$. The simplest examples in this class can be engineered by placing M2 and M5 branes as defects inside of a stack of background M5 branes. In this paper we show that a generalization of this construction yields more general solutions in the aforementioned class. The background branes are now M5's carrying M2 brane charge, while the defect branes are now placed at the origin of a flat hyperplane with a conical defect. The equations of motion imply a relation between the deficit angle produced by the conical defect and the M2 charge carried by the background branes.}

\begin{document}
\maketitle
\flushbottom

\section{Introduction}
Ever since the discovery of the AdS/CFT correspondence \cite{Maldacena:1997re} within the context of type IIB string theory, holographic methods have become an extremely valuable tool in the study of quantum field theories at strong coupling. This stimulated a massive production of supersymmetric supergravity backgrounds describing lower dimensional AdS vacua. Historically, the first examples where the correspondence was fully understood were $\mathrm{AdS}_5\times S^5$ \cite{Witten:1998qj}, $\mathrm{AdS}_4\times S^7$ \cite{Aharony:2008ug} and $\mathrm{AdS}_7\times S^4$, which are respectively dual to $\mathcal{N}=4$, $d=4$ SYM, 3d ABJM theories and $\mathcal{N}=(2,0)$ 6d SCFT's. In identifying and testing the correspondence a crucial role is played by the underlying brane picture, engineering the field theory on the one side of the correspondence, while yielding the dual AdS background on the other, upon taking the near-horizon limit.  

In the three aforementioned celebrated examples, the brane picture is simply given by a single stack of non-dilatonic branes, such as D3, M2 and M5 branes, respectively. The near-horizon geometry in these cases simply turns out to be given by the direct product of AdS times a sphere and no warping is present. When treating more involved situations featuring a non-trivial warping and a reduced amount of supersymmetry, it becomes increasingly difficult to obtain a complete understanding of supergravity AdS vacua in terms of brane systems. A collection of AdS warped geometries understood as near-horizon limits of brane intersections can be found e.g. in \cite{Cvetic:2000cj}.

From a pure supergravity viewpoint, an enormously helpful tool for exploring supersymmetric AdS solutions is represented by the pure spinor formalism \cite{Strominger:1986uh,Gauntlett:2002fz,Grana:2005sn}, which explicitly exploits the interplay between differential forms, spinors and geometry in order to determine and classify which geometries involving lower dimensional AdS factors may be consistently obtained as supersymmetric string theory backgrounds. However, despite triggering crucial developments in the production of novel fully-backreacted AdS warped geometries, this method is not particularly helpful when it comes to understanding the underlying brane system which originated them.
 
On the other hand, a physically well-understood mechanism that yields geometries of the above type is the backreaction of some defect branes inserted within the background of a pre-existing stack of branes. This phenomenon was originally studied in \cite{Karch:2000gx}, where this was proposed as the stringy description of defect CFT's. In this work, it was also shown, by means of a probe calculation, how the backreaction due to the presence of defect branes induces a warping on the worldvolume of the probe branes, which then becomes AdS times a sphere rather than flat Minkowski space.

If we focus on $\mathrm{AdS}_3$ solutions in particular, a wide range of possibilities opens up for what concerns the choice of internal manifolds, spin structures and preserved supersymmetries. Partial attempts of classification can be found e.g. in \cite{Argurio:2000tg,Kim:2005ez,Gauntlett:2006ns,Gauntlett:2006af,Donos:2008hd,DHoker:2008rje,Couzens:2017way,Eberhardt:2017uup,Dibitetto:2017tve,Dibitetto:2018ftj,Couzens:2019iog,Lozano:2019emq}.  Our focus in this paper is the class of solutions found in  \cite{Lozano:2020bxo}. The corresponding geometry is given by $\mrm{AdS}_3\times S^3/\mathbb{Z}_k\times \mrm{CY}_2\times I $ with warping w.r.t. the interval $I$, where special situations where $k=1$ and/or $ \mrm{CY}_2=\mathbb{R}^4$ describe cases with enhanced supersymmetry. In \cite{Lozano:2020bxo} in particular, two different branches of $\mrm{AdS}_3\times S^3/\mathbb{Z}_k\times \mrm{CY}_2\times I $ solutions are studied. In the former branch the warp factors in front of the AdS$_3$ and $S^3$ blocks of the metric are equal, while in the latter they differ by an extra factor. 

In \cite{Faedo:2020nol} the brane origin of the former branch of solutions is explained in terms of M2 -- M5 -- KK defects intersecting a stack of M5 -- KK branes. The aim of our work is to provide a brane set-up describing the other branch of solutions. The paper is organized as follows. In Sec. \ref{Sec2} we review the results of \cite{Faedo:2020nol}. Subsequently, in \ref{Sec3}, we introduce the novel ingredients needed in order to provide the brane interpretation of the other branch of AdS vacua. These will include M2 -- M5 defects at conical singulairities within the $ \mrm{CY}_2$, as well as exotic background five-branes carrying M2 charge. Finally, in \ref{Sec4} we present the full solution providing the brane interpretation for this second class of M-theory vacua.

\section{Review of M2 -- M5 -- M5 brane systems}
\label{Sec2}

In this section we summarize the results of \cite{Faedo:2020nol} where $\ma N=(0,4)$ $\mathrm{AdS}_3\times S^3/\mathbb{Z}_k\times {\tilde S}^3/\mathbb{Z}_{k'}\times \Sigma_2$ string backgrounds of M-theory have been obtained as near-horizon regime of ``defect" M2 -- M5 branes ending on a stack of orthogonal M5 branes with both the 5-branes placed at A-type singularities. These near-horizon geometries turn out to belong to a sub-class within the general classification of $\ma N=(0,4)$ $\mrm{AdS}_3\times S^3/\mathbb{Z}_k\times \mrm{CY}_2\times I $ solutions presented in \cite{Lozano:2020bxo} and thus provide, for this sub-class, a clear interpretation in terms of non-perturbative objects.

Let us consider a ``simplified" situation in which the KK-monopoles are absent since to the aims of this paper the inclusion of Taub-NUT singularities is not necessary. In this case the system we have to deal with preserves 8 supercharges and it is featured by M2 -- M5 branes completely localized within the worldvolume of the orthogonal M5 branes.
\begin{table}[http!]
\renewcommand{\arraystretch}{1}
\begin{center}
\scalebox{1}[1]{
\begin{tabular}{c||c c|c c c c | c c c c | c}
object & $t$ & $x^1$ & $r$ & $\theta^{1}$ & $\theta^{2}$ & $\theta^3$ &  $\rho$ & $\varphi^1$ & $\varphi^2$ & $\varphi^3$ & $z$     \\
\hline \hline
$\mrm{M}5$ & $\times$ & $\times$ & $\times$ & $\times$ & $\times$ & $\times$ & $-$ & $-$ & $-$ & $-$ & $-$  \\
\hline
$\mrm{M}2$& $\times$ & $\times$ & $-$ & $-$ & $-$ & $-$ &  $\sim$ & $\sim$ & $\sim$ & $\sim$ & $\times$  \\
$\mrm{M}5$ & $\times$ & $\times$ & $-$ & $-$ & $-$ & $-$ & $\times$ & $\times$ & $\times$ & $\times$ & $\sim$ \\
\end{tabular}
}
\end{center}
\caption{The $\frac14$ -- BPS brane system underlying the intersection of M2 -- M5 branes intersecting M5 branes. The ``$\sim$'' denotes smearing along the corresponding directions.} \label{Table:M2M5M5}
\end{table}
The metric for such a system has the form
\begin{equation}
\label{brane_metric_M2M5M5}
\begin{split}
d s_{11}^2&=H^{-1/3}\left(H_{\mathrm{M}5}^{-1/3}\,H_{\mathrm{M}2}^{-2/3}\,ds^2_{\mathrm{Mkw}_2}+H_{\mathrm{M}5}^{2/3}\,H_{\mathrm{M}2}^{1/3}\left(dr^2+r^2ds^2_{S^3}\right) \right)\\
&+H^{2/3}\left(H_{\mathrm{M}5}^{-1/3}\,H_{\mathrm{M}2}^{1/3}\,\left( d\rho^2+\rho^2ds^2_{\tilde S^3}\right)+H_{\mathrm{M}5}^{2/3}\,H_{\mathrm{M}2}^{-2/3}\,dz^2\right)\ .\\
\end{split}
\end{equation}
The smearing of M2 -- M5 brane charges implies that $H_{\mathrm{M}2}=H_{\mathrm{M}2}(r)$ and $H_{\mathrm{M}5}=H_{\mathrm{M}5}(r)$, while the orthogonal M5 branes are completely localized in their transverse space, i.e. $H=H(z,\rho)$. This charge distribution breaks the $\mathrm{SO}(3)\times\mathrm{SO}(3)$ invariance since, as we said, the M2 -- M5 charges are completely localized within the M5's worldvolume. The flux configuration corresponding to the aforementioned distribution of charges is given by
\begin{equation}
\begin{split}
\label{G4}
 G_{(4)}&=
 d\left( H_{\mathrm{M}2}^{-1}\right)\wedge\mathrm{vol}_{\mathrm{Mkw}_2}\wedge dz+ *_{(4)}\left(dH_{\mathrm{M}5}\right)\wedge dz\\
 &+\partial_\rho H\rho^3\, dz \wedge\text{vol}_{\tilde{S}^3}-H_{\mathrm{M}2}\,H_{\mathrm{M}5}^{-1}\,\partial_z H\rho^3\,d\rho\wedge\text{vol}_{\tilde{S}^3}\ ,
\end{split}
\end{equation}
where $*_{(4)}$ is performed within the $\mathbb R^4$ parametrized by $(r,\theta^i)$.
For this flux configuration, it turns out that the equations of motion and Bianchi identities of the defect branes and those of the orthogonal M5's decouple in the following way
\begin{equation}\label{M2M5EOM}
\begin{split}
& \left(r^3\dot{H}_{\mathrm{M}2}\right)^{\cdot}\,\overset{!}{=}\,0\ ,\qquad \Delta_{\mathbb{R}^4}\,H+\partial_z^2\,H\overset{!}{=}0\ ,
\end{split}
\end{equation}
where we denoted by $\cdot$ the derivative w.r.t. $r$ and $\Delta_{\mathbb{R}^4}$ represents the Laplace operator in the $\mathbb{R}^4$ space spanned by $\left(\rho,\varphi^{i}\right)$.
Moreover from the equations of motion for the 3-form one also infers that $H_{\mathrm{M}2}\overset{!}{=}H_{\mathrm{M}5}$.  The equation for $ H_{\mathrm{M}2}$ can be easily solved by
\begin{equation}\label{soldefectbranes}
  H_{\mathrm{M}2} \ = \ H_{\mathrm{M}5} \ = \ 1+\frac{Q_{\mathrm{M}2}}{r^2}\ ,
 \end{equation}
 where $Q_{\mathrm{M}2}$ are the charges of M2 (M5) branes.
 
Let us now consider the  $r\rightarrow 0$ limit. In this regime the full 11d background takes the form of a stack of M5 branes wrapping an $\mrm{AdS}_3\times S^3$ geometry of M2 -- M5 branes\footnote{We redefined the 2d Minkowski coordinates as $(t,x^1)\rightarrow Q_{\mathrm{M}2}\,(t,x^1)$.}
\begin{equation}
\label{nh_brane_metric_M2M5M5}
\begin{split}
d s_{11}^2&=Q_{\text{M}2}\,H^{-1/3}\,\left(ds^2_{{\scriptsize \mrm{AdS}_3}}+ds^2_{S^3} \right)+H^{2/3}\left(ds^2_{\mathbb{R}^4}+dz^2\right) \, ,\\
 G_{(4)}&=2 \,Q_{\text{M}2}\,\text{vol}_{{\scriptsize \mrm{AdS}_3}}\wedge dz+2\,Q_{\text{M}2}\,\text{vol}_{S^3}\wedge dz\\
 &-\partial_z H\rho^3\,d\rho\wedge \text{vol}_{\tilde{S}^3}
 +\partial_\rho H\rho^3\,dz \wedge \text{vol}_{\tilde{S}^3}\ ,
\end{split}
\end{equation}
where $ds^2_{\mathbb{R}^4}=d\rho^2+\rho^2\,ds^2_{\tilde S^3}$ and the function $H$ is a harmonic function on the 5d flat transverse space of the M5 branes parametrized by $(\rho,z)$.
In this regime the solution enhances its supersymmetry to 16 supercharges preserved ($\frac12$ -- BPS). In particular it realizes small $\ma N=(4,4)$ supersymmetry since only one of the two $\mrm{SU}(2)$ symmetry factors is preserved.

\subsection*{The inclusion of KK monopoles}

At the level of the local description of near-horizon background \eqref{nh_brane_metric_M2M5M5}, the possible inclusion of KK-monopoles is technically straightforward since it consists in the mere substitution of the 3-sphere\footnote{The presence of a Taub-NUT singularity within the flat space parametrized by $(r,\theta^i)$ allows a further inclusion of an A-type singularity within the orthogonal $\mathbb{R}^4$ parametrized by $(\rho,\phi^i)$ without any further breaking of SUSY \cite{Faedo:2020nol}.} appearing in \eqref{nh_brane_metric_M2M5M5} with a lens-space, i.e. $S^3\rightarrow S^3/\mathbb{Z}_k$. This operation implies a further breaking of a quarter of supersymmetry and it produces $\mrm{AdS}_3$ backgrounds preserving small $\ma N=(0,4)$ supersymmetry \cite{Faedo:2020nol}. These backgrounds are included in the classification of $\ma N=(0,4)$ $\mrm{AdS}_3\times S^3/\mathbb{Z}_k\times \mrm{CY}_2\times I $ solutions studied in \cite{Lozano:2020bxo}.
\begin{table}[http!]
\renewcommand{\arraystretch}{1}
\begin{center}
\scalebox{1}[1]{
\begin{tabular}{c||c c|c c c c | c c c c | c  }
object & $t$ & $x^1$ & $r$ & $\theta^{1}$ & $\theta^{2}$ & $y$  & $\rho$ & $\varphi^1$ & $\varphi^2$ & $\phi$ & $z$    \\
\hline \hline
$\mrm{M}5$ & $\times$ & $\times$ & $\times$ & $\times$ & $\times$ & $\times$ & $-$ & $-$ & $-$ & $-$ & $-$ \\
\hline
$\mrm{M}2$& $\times$ & $\times$ & $-$ & $-$ & $-$ & $-$  & $\sim$ & $\sim$ & $\sim$ & $\sim$ & $\times$ \\
$\mrm{M}5$ & $\times$ & $\times$ & $-$ & $-$ & $-$ & $-$ & $\times$ & $\times$ & $\times$ & $\times$ & $\sim$\\
$\mrm{KK}$ & $\times$ & $\times$ &$-$ & $-$ & $-$ & $\text{ISO}$ & $\times$ & $\times$ & $\times$  & $\times$ & $\times$\\
\end{tabular}
}
\end{center}
\caption{$\frac18$ -- BPS brane setup of defect M2 -- M5 branes intersecting orthogonal M5 branes with KK monopoles. $y$ is the compact direction of the KK monopoles \cite{Faedo:2020nol}.} \label{Table:M2M5M5KK}
\end{table}
At the level of the brane picture the above arguments are realized by including KK-monopoles charges $Q_{\mathrm{KK}}$ in the transverse spaces of the M2 -- M5 defect branes. The full brane solution has been studied in \cite{Faedo:2020nol} and here we just present the metric
\begin{equation}
\label{metric_M2M5M5KKmopoles}
\begin{split}
d s_{11}^2&=H^{-1/3}\left[H_{\mathrm{M}5}^{-1/3}H_{\mathrm{M}2}^{-2/3}ds^2_{\mathrm{Mkw}_2}+H_{\mathrm{M}5}^{2/3}H_{\mathrm{M}2}^{1/3}\left(H_{\text{KK}}ds^2_{\mathbb{R}^3}+H_{\text{KK}}^{-1}(dy+Q_{\mathrm{KK}}\omega)^2\right) \right]\\
&+H^{2/3}\left[H_{\mathrm{M}5}^{2/3}H_{\mathrm{M}2}^{-2/3}dz^2+H_{\mathrm{M}5}^{-1/3}H_{\mathrm{M}2}^{1/3}ds^2_{\mathbb{R}^4}\right]\ ,\\
\end{split}
\end{equation}
with $d\omega=\text{vol}_{S^2}$ and $ds^2_{\mathbb{R}^3}=dr^2+r^2ds^2_{S^2}$. The parameter $Q_{\mathrm{KK}}$ coincides with the $\mathbb{Z}_k$-orbifold of the 3-sphere, namely $k=Q_{\mathrm{KK}}$. This can be showed explicitly
by changing the radial coordinates as $r \rightarrow 4^{-1}\,Q_{\text{KK}}^{-1}\,r^2$ \cite{Cvetic:2000cj}, in this way the 4d flat manifold parametrized by $(r, \theta^i, y)$ can be written as foliations of the Lens space $S^3/\mathbb{Z}_{k}$.

Finally it worths to mention that in \cite{Faedo:2020nol} it has been showed that the aforementioned $\ma N=(0,4)$ $\mrm{AdS}_3$ backgrounds reproduce asymptotically a (locally) $\mrm{AdS}_7/\mathbb{Z}_k\times  S^4$ geometry, allowing an interpretation in holography as a $\ma N=(0,4)$ conformal surface defect. In what follows we will always consider M-branes intersections that do not include KK monopoles, but that allows their inclusion as an immediate generalization following the aformentioned prescriptions.

\section{The new ingredients: the M5$_{\xi}$ brane and conical defects}
\label{Sec3}

We have seen in the previous section that a specific type of AdS$_3$ solutions to 11d supergravity can be interpreted as the near-horizon geometries of M2 -- M5 defect branes placed within a background given by a stack of M5 branes. With  the inclusion of KK monopoles, the aforementioned solutions turn out to be a subclass of $\ma N=(0,4)$ $\mrm{AdS}_3\times S^3/\mathbb{Z}_k\times \mrm{CY}_2\times I $ backgrounds studied in \cite{Lozano:2020bxo}.

The goal of this section is to discuss the new ingredients needed to find the brane picture of more general AdS$_3$ vacua within the class of \cite{Lozano:2020bxo}, namely $\mrm{AdS}_3\times S^3$ sliced backgrounds described by a relative warping\footnote{In the notation of \cite{Lozano:2020bxo} this additional warping is encoded in a linear function called $u$.} between $\mrm{AdS}_3$ and $S^3$.
To this aim we will need to deform both the background objects (the M5 branes) and the defect ones (the M2 -- M5 bound state). In this section we will start out by analyzing these two deformations separately and eventually, in the next section, we shall present the complete solution in which these two effects are combined. As an effect of superimposing the two, a dynamical constraint will finally relate the parameters controlling the two deformations.

\subsection*{The M5$_{\xi}$ brane solution}

In addition to M-branes and their intersections, another class of objects feature the spectrum of M-theory.
The first example of these objects was first found in \cite{Izquierdo:1995ms} by studying the compactification of M-theory on a $T^3$. In this work a ``dyonic" multi-membrane solution was worked out within $\ma N=2$ 8d supergravity by exploiting the $\mrm{SL}(2,\mathbb{R})$ invariance of the theory. This solution represents
a solitonic five-brane carrying membrane charge distributed within a 3d hypersurface inside of its worldvolume.

From the point of view of the physics of M-branes, this is equivalent to considering an exotic (M2, M5) bound state given by an M5 brane carrying a dissolved M2 charge within its worldvolume. This object preserves 16 supercharges in 11d.

The explicit form of the solution is specified by the choice of a harmonic function $H$ over the transverse space $\mathbb{R}^5$, and a duality angle $\xi$ fixing the ratio between the (magnetic) five-brane charge and the (electric) membrane charge deforming its worldvolume. We will refer to this object as M5$_{\xi}$ or \emph{dyonic membrane}.
%
%
\begin{table}[http!]
\renewcommand{\arraystretch}{1}
\begin{center}
\scalebox{1}[1]{
\begin{tabular}{c||ccc|ccc|ccccc}
object & $t$ & $x^1$ & $x^2$ & $u^1$ & $u^2$ & $u^3$ &  $\zeta$ & $\theta^1$ & $\theta^2$ & $\theta^3$ & $\theta^4$     \\
\hline \hline
$\mathrm{M5}_{\xi}$ & $\times$ & $\times$ & $\times$ & $\times$ & $\times$ & $\times$ & $-$ & $-$ & $-$ & $-$ & $-$ \\
\end{tabular}
}
\end{center}
\caption{The M5$_{\xi}$ or dyonic membrane describing M5 branes with M2 charges dissolved along the hypersurface parametrized by $(u^1,u^2,u^3)$. The transverse space $\mathbb{R}^5$ is parametrized by the coordinates $(\zeta, \theta^1, \theta^2,\theta^3,\theta^4 )$. The object is $\frac12$ -- BPS.} \label{Table:Lambert}
\end{table}
The explicit form of the solution is given by \cite{Izquierdo:1995ms}
\be
\begin{array}{lclc}\label{lamberbrane}
ds_{11}^{2} & = & H^{-2/3}\,\left(s^2+c^2H\right)^{1/3}ds_{\mathrm{Mkw}_3}^2 +H^{1/3}\,\left(s^2+c^2H\right)^{-2/3}ds_{\mathbb{R}^3}^2+\\ 
& & H^{1/3}\,\left(s^2+c^2H\right)^{1/3}\,\left( d\zeta^2+\zeta^2ds_{S^4}^2 \right) & , \\[2mm]
G_{(4)} & = & c*_{(5)}(dH)+  s\,d\left(H^{-1}\right)\wedge\mathrm{vol}_{\mathrm{Mkw}_3} + \dfrac{sc}{\left(s^2+c^2H\right)^{2}}dH\wedge\mathrm{vol}_{\mathbb{R}^3}& ,
\end{array}
\ee
where $s\,\equiv\,\sin\xi$ and $c\,\equiv\,\cos\xi$, whereas $H$ satisfies $\Delta_{\mathbb{R}^5}H=0$. We immediately observe that \eqref{lamberbrane} admits two relevant limits:

\begin{itemize}
 \item $\xi=0$: Purely magnetic case describing fully-localized M5 branes.
 
 \item $\xi=\frac{\pi}{2}$: Purely electric case describing M2 branes smeared over the hyperplane spanned by $(u^1, u^2,u^3)$.
\end{itemize}
For any other value of the duality angle $\xi$, one has an exotic five-brane carrying membrane charge. We point out that the presence of the last term in $G_{(4)}$ makes it manifest that we are not dealing with a mere superposition of M2 and M5 branes. In fact such an intersection would drastically break all of the supersymmetries, while the presence of such a term in $G_{(4)}$ crucially saves us from SUSY breaking by taking M2 -- M5 interactions  into account.

A particular asymptotically flat solution is the one describing an object carrying spherically symmetric charge in the transverse space $\mathbb{R}^5$, which corresponds to choosing $H$ as 
\be
\label{Harm_H}
H \ = \ 1\,+\,\frac{Q}{\zeta^{3}} \ ,
\ee
where $Q$ represents the five-brane charge. It is perhaps worth noticing that, for any $\xi\neq\frac{\pi}{2}$, the near-horizon geometry of this object will still yield an AdS$_7$ factor. To see this explicitly, take the $\zeta\rightarrow 0$ limit to get a metric of the form
\be
\label{AdS7_limit}
\ell^{-2}\,ds_{11}^2\,\sim\, \left(\frac{d\zeta^{2}}{\zeta^{2}}\,+\,\zeta\,\left(\frac{c^{1/3}}{Q}\,ds_{\mathrm{Mkw}_3}^2+\frac{c^{-1}}{Q}\,ds_{\mathbb{R}^3}^2\right)\right) \, + \, ds_{S^4}^2 \ ,
\ee
with $\ell\equiv c^{1/6}Q^{1/3}$, which can be recognized as the direct product of $\mathrm{AdS}_7$ and $S^4$, after a suitable rescaling of the coordinates of $\mathrm{Mkw}_3$ and $\mathbb{R}^3$, respectively.

\subsection*{M2 -- M5 intersections at conical defects}

Let us now have a look at the deformation needed for an M2 -- M5 defect to be compatible with an $\mathrm{M5}_{\xi}$ background. First of all we start by considering the M2 -- M5 set-up included in \eqref{brane_metric_M2M5M5}, where the orthogonal M5 branes have been removed (see Table \ref{Table:M2M5M5conical}). The common transverse space to the defect will be an $\mathbb{R}^4$ spanned by the polar coordiantes $\left(r,\theta^i\right)$ and the M2 and M5 branes are fully localized in this space. This time though, a \emph{conical defect} must be placed at the origin of this hyperplane, i.e. at $r=0$. Let us first see how this conical defect is introduced within $\mathbb{R}^4$.

By adopting embedding coordinates in $\mathbb{R}^5$, call them $\left\{X^{I}\right\}$, the embedding of a four-dimensional cone is defined through
\be
\left(X^{5}\right)^{2} \, - \, \frac{\left|1-\gamma^2\right|}{\gamma^2} \, \left(X^{a}X^{a}\right) \, = \, 0 \ ,
\ee
where $\gamma$ is a non-zero constant and $a$ runs from 1 to 4. Upon introducing 4d local coordinates through
\be
\begin{array}{lccclc}
X^{a} \, = \, \gamma \, r\mu^{a} & , & & &  X^{5} \, = \, \left|1-\gamma^2\right|^{1/2}\,r & ,
\end{array}
\ee
where $\left\{\mu^a\right\}$ are the embedding coordinates of a three-sphere and hence satisfy $\mu^a\mu^a = 1$. The induced metric then reads
\be
ds_{\mathbb{R}^4_{\gamma}}^2 \, = \, dr^2 \, + \, \gamma^2 \, r^2 ds_{S^3}^2 \ ,
\ee
which corresponds to a hyperplane with deficit angle $\delta\equiv\,2\pi\,(1-\gamma)$ and it contains a conical singularity at $r=0$ for any $\gamma\neq 1$. Note that the apex angle $2\psi$ of the corresponding cone is related to the $\gamma$ parameter through $\gamma=\sin\psi$.

Let us now place our M2 -- M5 defect branes as depicted in Table \ref{Table:M2M5M5conical}, where the M5 charge is uniformly distributed along the coordinate $z$ and the M2 branes are smeared over the worldvolume of the M5.
\begin{table}[http!]
\renewcommand{\arraystretch}{1}
\begin{center}
\scalebox{1}[1]{
\begin{tabular}{c||c c|c c c c | c c c c | c}
object & $t$ & $x^1$ & $r$ & $\theta^{1}$ & $\theta^{2}$ & $\theta^3$ &  $\rho$ & $\varphi^1$ & $\varphi^2$ & $\varphi^3$ & $z$     \\
\hline \hline
$\mrm{M}2$& $\times$ & $\times$ & $-$ & $-$ & $-$ & $-$ &  $\sim$ & $\sim$ & $\sim$ & $\sim$ & $\times$  \\
$\mrm{M}5$ & $\times$ & $\times$ & $-$ & $-$ & $-$ & $-$ & $\times$ & $\times$ & $\times$ & $\times$ & $\sim$ \\
\end{tabular}
}
\end{center}
\caption{The $\frac14$ -- BPS brane system underlying the intersection of M2 -- M5 branes intersecting M5 branes. The conical singularity is located within the $\mathbb{R}^4_{\gamma}$ parametrized by $(r,\theta^i)$.} \label{Table:M2M5M5conical}
\end{table}

%
 The effect of placing a conical singularity at the origin of the $\mathbb{R}^4$ parametrized by $\left(r,\theta^i\right)$ is a backreaction that appears in the metric through a $z$ dependent factor named $f$.
The explicit solution is given by
\be
\begin{array}{lclc}\label{M2M5conical}
ds_{11}^{2} & = & f^{2/3}\,\left(H_{\mathrm{M}2}^{-2/3}H_{\mathrm{M}5}^{-1/3}ds_{\mathrm{Mkw}_2}^2\,+\,H_{\mathrm{M}2}^{1/3}H_{\mathrm{M}5}^{2/3}\left(dr^2+\gamma^2r^2ds^2_{S^3} \right)\right) +\\[1.5mm]
&+&  f^{-1/3}\left( H_{\mathrm{M}2}^{1/3}H_{\mathrm{M}5}^{-1/3} \left(d\rho^2+\rho^2ds^2_{\tilde S^3} \right) + f^{-1} H_{\mathrm{M}2}^{-2/3}H_{\mathrm{M}5}^{2/3} dz^2\right) & , \\[2mm]
G_{(4)} & = &\gamma^{-1} d\left( H_{\mathrm{M}2}^{-1}\right)\wedge\mathrm{vol}_{\mathrm{Mkw}_2}\wedge dz+ *_{(4)}\left(dH_{\mathrm{M}5}\right)\wedge dz \,.&
\end{array}
\ee
where $*_{(4)}$ is performed over $\mathbb R^4_\gamma$.
From the equations of motion and Bianchi it follows that the functions $H_{\mathrm{M}2}(r)$ and $H_{\mathrm{M}5}(r)$ both satisfy the harmonicity condition on $\mathbb{R}^4_\gamma$, while $f$ has to be a linear function of $z$, 
\begin{equation}
\left(r^3\dot{H}_{\alpha}\right)^{\cdot}\,\overset{!}{=}\,0 \quad \text{with} \,, \qquad  f''\,\overset{!}{=}\,0,
\end{equation}
where $\alpha=\mathrm{M}2,\,\mathrm{M}5$ and we denoted respectively by $\cdot\,\,$ $\&$ $\,\,\prime$ the derivatives with respect to $r$ $\&$ $z$.
Moreover, the full set of EOM's and Bianchi identities require the absence of constant term inside  $H_{\mathrm{M}2}(r)$ and fix $f'$ in terms of $\gamma$. Summarizing, the complete solution for the defect is gven by
\be
\begin{array}{lclc}\label{solM2M5defect}
 H_{\mathrm{M}2}\, =\, \dfrac{Q_{\mathrm{M}2}}{r^2}  & , &\qquad  H_{\mathrm{M}5} \, = \, c_{\mathrm{M}5} \, + \,  \dfrac{Q_{\mathrm{M}5}}{r^2} & ,\\[3mm]
   f \, = \, c \, + \, Qz & ,& \qquad \text{with} \qquad  Q\,=\,\pm\frac{2}{\sqrt{Q_{\mathrm{M}2}}}\,\frac{\sqrt{1-\gamma^2}}{\gamma} & ,
\end{array}
\ee
where 
 $c_{\mathrm{M}5}$ and $c$ are arbitrary constants. The corresponding geometry when zooming in at $r\rightarrow 0$ is given by\footnote{We redefined the 2d Minkowski coordinates as $(t,x^1)\rightarrow Q_{\mathrm{M}2}^{1/2}Q_{\mathrm{M}5}^{1/2}\,(t,x^1)$.},
\be
\begin{array}{lclc}
ds_{11}^{2} & = & Q_{\mathrm{M}2}^{1/3}Q_{\mathrm{M}5}^{2/3}\,f^{2/3}\,\left(ds^2_{\mathrm{AdS}_3}+\gamma^2 ds^2_{S^3}\right) \,+\,  f^{-1/3}ds_{\mathbb{R}^4}^2 + f^{-4/3}  dz^2 & , \\[2mm]
G_{(4)} & = & 2Q_{\text{M}5}\gamma^{-1}\,\text{vol}_{{\scriptsize \mrm{AdS}_3}}\wedge dz+2\,Q_{\text{M}5}\gamma^{3}\,\text{vol}_{S^3}\wedge dz ,&
\end{array}
\ee
where $ds_{\mathbb{R}^4}^2=d\rho^2+\rho^2ds^2_{\tilde S^3}$.
This near-horizon metric constitutes a $\mathrm{AdS}_{3}\times S^3\times \mathbb{R}^4$ solution of M-theory with warping along the 11th $z$ coordinate. This background preserves the same number of supersymmetries of the M2 -- M5 set-up without any conical singularity. This is obviously related to the presence of the linear function $f$ whose effect is exactly that of compensating the deformation due to the conical singularity.

\section{The full solution and its limits}
\label{Sec4}

After analyzing the background and defect branes separately, we are now ready to study the complete brane system realizing the general $\mathrm{AdS}_3$ solutions in \cite{Lozano:2020bxo}. The underlying brane picture is sketched in table~\ref{Table:Mbranes} and it describes a stacks of dyonic membranes M5$_\xi$ orthogonal to M2 -- M5 defect branes at a conical singularity. The crucial element we observe is the impossiblity for the dyonic membrane to wrap the defect geometry of M2 -- M5 unless one does not include a conical singularity. This is related to the fact that the M2 charge extended within the worldvolume of dyonic membranes breaks the isometries of its 6d worldvolume and this effect has to be compensated by the backreacted geometry of the defect branes M2 -- M5.
\begin{table}[h!]
\renewcommand{\arraystretch}{1}
\begin{center}
\scalebox{1}[1]{
\begin{tabular}{c||c c|c c c c|c c c c |c}
branes & $t$ & $x$  & $r$ & $\theta^{1}$ & $\theta^{2}$ & $\theta^{3}$ & $\rho$ & $\varphi^{1}$ & $\varphi^{2}$ & $\varphi^{3}$ & $z$ \\
\hline \hline
M5$_{\xi}$ & $\times$ & $\times$  & $\times$ & $\times$ & $\times$ & $\times$ & $-$ & $-$ & $-$ & $-$ & $-$  \\
\hline
M2 & $\times$ & $\times$ & $-$ & $-$ & $-$ & $-$ & $\sim$ & $\sim$ & $\sim$ & $\sim$ & $\times$ \\
M5 & $\times$ & $\times$ & $-$ & $-$ & $-$ & $-$ & $\times$ & $\times$ & $\times$ & $\times$ & $\sim$ \\
\end{tabular}
}
\end{center}
\caption{{\it The brane picture underlying the 2d SCFT described by M2 and M5 branes as defects within the 6d $\mathcal{N}=(2,0)$ SCFT living on the worldvolume of the M5$_{\xi}$. The above system is $\frac{1}{4}$ -- BPS.
}} \label{Table:Mbranes}
\end{table}
Based on the previous analysis, we can cast the following \emph{Ansatz} for the 11d metric
\be
\begin{array}{lclc}\label{lambertdefect}
ds_{11}^{2} & = &  f^{2/3}\,H^{-2/3}\,\left(s^2+c^2H\right)^{1/3}\left(H_{\mathrm{M}2}^{-2/3}H_{\mathrm{M}5}^{-1/3}ds_{\mathrm{Mkw}_2}^2+H_{\mathrm{M}2}^{1/3}H_{\mathrm{M}5}^{2/3}dr^2\right) +\\[1.5mm]
&+& f^{2/3}\,H^{1/3}\,\left(s^2+c^2H\right)^{-2/3}\,H_{\mathrm{M}2}^{1/3}H_{\mathrm{M}5}^{2/3}\,\gamma^2 r^2 ds_{S^3}^2+\\ [1.5mm]
&+ & f^{-1/3}H^{1/3}\,\left(s^2+c^2H\right)^{1/3}\,\left( H_{\mathrm{M}2}^{1/3}H_{\mathrm{M}5}^{-1/3}\left ( d\rho^2+\rho^2ds^2_{\tilde S^3}\right) + f^{-1} H_{\mathrm{M}2}^{-2/3}H_{\mathrm{M}5}^{2/3} dz^2\right) & , \\
\end{array}
\ee
where the function $H(\rho,z)$ is now associated with the dyonic membrane M5$_\xi$ and $s=\sin\xi$, $c=\cos \xi$ as they have been introduced in \eqref{lamberbrane}. The functions $H_{\mathrm{M}2}(r)$ and $H_{\mathrm{M}5}(r)$ describe the defect branes M2 -- M5 at the conical singularity parametrized by the parameter $\gamma$. Finally the function $f(z)$ is the additional warping appearing when placing the defect at the singularity as in \eqref{M2M5conical}. The corresponding 4-flux has the following form
\be
\begin{array}{lclc}
G_{(4)} & = & c\, d\left( H_{\mathrm{M}2}^{-1}\right)\wedge\mathrm{vol}_{\mathrm{Mkw}_2}\wedge dz+c^3 *_{(4)}\left(dH_{\mathrm{M}5}\right)\wedge dz + \\[1.5mm]
&-&s\,H_{\mathrm{M}2}^{-1/2}\mathrm{vol}_{\mathrm{Mkw}_2}\wedge dr\wedge d\left(fH^{-1}\right) -sc^2\,r^3H_{\mathrm{M}5}^{3/2}\mathrm{vol}_{S^3} \wedge d\left(f\left(s^2+c^2H\right)^{-1}\right)+\\[1.5mm]
& +& c\rho^3\left(f^{-1}\partial_{\rho}Hdz-H_{\mathrm{M}2}\,H_{\mathrm{M}5}^{-1}\partial_z Hd\rho\right)\wedge\, \mathrm{vol}_{\tilde{S}^3} \, ,& 
\end{array}
\ee
where $*_{(4)}$ is performed over $\mathbb R^4_\gamma$.
By looking at the above expression we can recognize in the first two terms the contribution of defect branes M2 -- M5 written in \eqref{M2M5conical}, while the other terms represent deformations of the 4-flux of the dyonic membrane \eqref{lamberbrane} due to the backreaction of the conical singularity.

The functions $H(\rho,z)$, $f(z)$, $H_{\mathrm{M}2}(r)$ and $H_{\mathrm{M}5}(r)$ satisfy the following system of differential conditions
\be
\begin{array}{lclclclc}\label{EOMdefectlambert}
\left(r^3\dot{H}_{\mathrm{M}2}\right)^{\cdot}\,\overset{!}{=}\,0\ , &  & \left(r^3\dot{H}_{\mathrm{M}5}\right)^{\cdot}\,\overset{!}{=}\,0 & , \\[1.5mm]
f^{-1}\Delta_{\mathbb{R}^4}H+\partial_z^2H\overset{!}{=}\,0 & , &\qquad f''\overset{!}{=}\,0 \ ,
\end{array}
\ee
where $\cdot$ \& $\prime$ again denote differentiation w.r.t. $r$ \& $z$, respectively, and $\Delta_{\mathbb{R}^4}$ represents the Laplace operator in the $\mathbb{R}^4$ space spanned by $\left(\rho,\varphi^{i}\right)$. If one compares this situation with the case of \eqref{M2M5EOM}, the equation of $H$ does not decouple completely from the defect since it is explicitly deformed by the function $f$ associated to the conical singularity.

As in the case of \eqref{M2M5EOM}, the equation of motion of the 3-form $A_{(3)}$ turns out to further demand $H_{\mathrm{M}2}\,\overset{!}{=}\,H_{\mathrm{M}5}$, thus resulting in the absence of constant term in \eqref{solM2M5defect} for $H_{\mathrm{M}5}$, as well.
Crucially, one obtains a dynamical relation between the characteristic parameter of the conical defect $\gamma$ and the duality angle $\xi$ of the M5$_{\xi}$ brane:
\be
\gamma \ \overset{!}{=} \ c \ \equiv \ \cos\xi \ , 
\label{xi2gamma_relation}
\ee
which in turn implies that the duality angle $\xi$ needs to be the complement to $\frac{\pi}{2}$ of the half apex angle of the cone.
Finally, due to \eqref{xi2gamma_relation}, the first derivative of the $f$ function can now be directly expressed in terms of $\xi$ through
\be
Q  \ \overset{!}{=} \ -\frac{2}{\sqrt{Q_{\mathrm{M}2}}}\,\tan\xi \ .
\label{xi2Q_relation}
\ee
Hence, as for the M2 -- M5 system at a conical singularity \eqref{solM2M5defect}, $f^\prime$ is again fixed in terms of $\gamma$. We can then see that the inclusion of a conical singularity within the transverse space of defect branes M2 -- M5 is the necessary condition for their intersection with the dyonic membrane M5$_\xi$. In fact, as we will discuss in more detail in next section, one can immediately see from \eqref{xi2Q_relation} that the angle $\xi$ is trivialized when ones set $f^\prime=Q=0$ ane the 11d metric \eqref{lambertdefect} collapses to the already-knwon case of M2 -- M5 ending on a stack of M5 branes studied in section \ref{Sec2}.

\subsection*{Recognizing the underlying AdS$_3$ solution}

 Once explicitated all the quantities regarding the defect branes M2 --M5, the explicit background \eqref{lambertdefect} reproduces the dyonic membrane wrapping the fully-backreacted geometry of the defect,
\be
\label{AdS3_S3_f}
\begin{array}{lclc}
ds_{11}^{2} & = &  f^{2/3}\,\left(H^{-2/3}\left(s^2+c^2H\right)^{1/3}L^2\,ds_{\mathrm{AdS}_3}^2+ \,H^{1/3}\left(s^2+c^2H\right)^{-2/3} \kappa^2\,ds_{S^3}^2\right)\,+\\[1mm]
& + &  f^{-1/3}H^{1/3}\,\left(s^2+c^2H\right)^{1/3}\,\left(d\rho^2+\rho^2\,ds_{\tilde{S}^3}^2 + f^{-1} \, dz^2\right) & , \\[3mm]
G_{(4)} & = & -s\,L^3\,\mathrm{vol}_{\mathrm{AdS}_3}\wedge d\left(f\,H^{-1}\right) -sc^{-1}\,\kappa^3\,\mathrm{vol}_{S^3} \wedge d\left(f\,\left(s^2+c^2H\right)^{-1}\right)\,+\\[1mm]
&+&\left(Q_{\mathrm{M}2}L^3 \mathrm{vol}_{\mathrm{AdS}_3}+Q_{\mathrm{M}5}\kappa^3 \mathrm{vol}_{S^3} \right) \wedge dz \, + \, c\rho^3\left(f^{-1}\partial_{\rho}Hdz-\partial_z Hd\rho\right)\wedge\, \mathrm{vol}_{\tilde{S}^3} & ,
\end{array}
\ee
where the field equations and the Bianchi identities fix the ratio of the defect charges\footnote{To avoid ambiguities with the solutions presented in previous sections, we point out that the defect charges $Q_{\mathrm{M5}}$ and $Q_{\mathrm{M2}}$ we have chosen to present the $\mrm{AdS}_3$ solution \eqref{AdS3_S3_f} do not coincide with the parameters used in previous section for the corresponding full brane solution \eqref{lambertdefect}.} in terms of the duality angle $\xi$ as
\be
\frac{Q_{\mathrm{M}2}}{Q_{\mathrm{M5}}} \ \overset{!}{=} \ -\,c^{3} \ ,
\ee
while the AdS and the sphere radii must be given by
\be
\begin{array}{lccclc}
L \ = \ \dfrac{2}{Q_{\mathrm{M5}}}\,c^{-2} & , & & & \kappa \ = \  \dfrac{2}{Q_{\mathrm{M5}}}\,c^{-1} & ,
\end{array}
\ee
and finally the first derivative of the linear function $f$ reads $Q\,=\, Q_{\mathrm{M5}} \, sc$.

Note that, as already anticipated since the very beginning, this construction yields an $\mathrm{AdS}_{3}\times S^{3}\times \tilde{S}^{3}\times\Sigma_{2}$  solution in which the relative warping between $\mathrm{AdS}_{3}$ and the first $S^3$ are different, while in the construction review in section \ref{Sec2}, one always gets equal warp factors for the two. Note that, due to the absence of extra KK monopoles, these solutions preserve sixteen real supercharges and hence can be fit within the classification of \cite{DHoker:2008rje}.
The special situation with no relative warping between  $\mathrm{AdS}_{3}$ and $S^{3}$ can be recovered by taking the $\xi\rightarrow 0$ limit. It is worth mentioning that, within this parametrization, one smoothly approaches the solution \eqref{brane_metric_M2M5M5} as $\xi\rightarrow 0$. Taking the duality angle to zero, by taking the on-shell relations \eqref{xi2gamma_relation} \& \eqref{xi2Q_relation}, physically identifies the following limiting procedure
\begin{itemize}
\item M5$_{\xi} \ \rightarrow \ $M5, i.e. the background M2 charge disappears leaving us with a plain M5 brane,
\item $\gamma \ \rightarrow \ 1$, i.e. the conincal singularity in $\mathbb{R}^4$ disappears,
\item $f' \ \rightarrow \ 0$, i.e. no more extra warping in the $x^{10} \,=\,z$ coordinate.
\end{itemize}
These are all physical features of the $\mathrm{AdS}_{3}\times S^{3}\times \tilde{S}^{3}\times\Sigma_{2}$ solution reviewed in section \ref{Sec2}, which can be obtained as the near-horizon limit of an M2 -- M5 -- M5 brane system.

\subsection*{Recovering the AdS$_7$ asymptotic geometry}

Let us now take a closer look at the second order PDE satisfied by the function $H(\rho,z)$
\be
f^{-1}\Delta_{\mathbb{R}^4_{\rho}}H+\partial_z^2H\,=\,0 \ .
\label{PDE_H}
\ee
Since we are describing M2 -- M5 defect branes placed inside of a background M5$_{\xi}$ brane, we would expect to be able to recover an (asymptotically locally) AdS$_7\times S^4$ geometry when approaching the origin of the 5d transverse space of the M5$_\xi$ parametrized by $(\rho,z)$, just like it is discussed right above equation \eqref{AdS7_limit}. However, one might at first sight be worried by the fact that the solution to the PDE in \eqref{PDE_H} is represented by a \emph{non-harmonic} function, contrary to the one spelled out in \eqref{Harm_H}. Nevertheless, even though writing down a full closed-form solution for $H$ is not an easy task, one gets easily convinced that, at least in the $(\rho,z)\rightarrow (0,0)$ limit, the Laurent expansion for $H$ will be dominated by the harmonic function $H_{0}$ solving
\be
f(0)^{-1}\Delta_{\mathbb{R}^4_{\rho}}H_0+\partial_z^2H_0\,=\,0 \ .
\ee
By further redefining things s.t. $f(0)=1$, $H_0$ is just given by the expression  \eqref{Harm_H}, where $\zeta\,\equiv\,\left(\rho^2+z^2\right)^{1/2}$. Hence, when approaching the origin, the metric in \eqref{AdS3_S3_f} exactly takes the form of \eqref{AdS7_limit}, but with $ds_{\mathrm{Mkw}_3}^2$ \& $ds_{\mathbb{R}^3}^2$ replaced by $ds_{\mathrm{AdS}_3}^2$ \& $ds_{S^3}^2$, respectively. This defines an $\mathrm{AdS}_3\times S^3$ foliation of an asymptotically locally $\mathrm{AdS}_7$ geometry over the $\lambda\,\equiv\,\log\zeta$ coordinate. Deviations from empty $\mathrm{AdS}_7$ are subleading in $\lambda$, and are supported by the presence of non-vanishing components of $A_{(3)}$ wrapping both vol$_{ \mathrm{AdS}_3}$ and vol$_{S^3}$, both yielding non-vanishing $G_{(4)}$ flux.

\subsection*{Inclusion of KK monopoles and $\ma N=(0,4)$ solutions}

The solutions $\mrm{AdS}_3\times S^3 \times \tilde S^3 \times \Sigma_2$ presented in \eqref{AdS3_S3_f} preserves 16 supercharges. Let's shortly discuss the inclusion of KK monopoles to produce $\ma N=(0,4)$ solutions. As we exlained in section \ref{Sec2}, from the point of view of the local supergravity picture the inclusion of a Taub-NUT singularity is equivalent, to the mere substitution of $S^3\rightarrow S^3/\mathbb{Z}_k$ in the background \eqref{AdS3_S3_f}. We refer to section \ref{Sec2} where the prescriptions to include KK monopoles to these types of backgrounds are outlined.
\begin{table}[http!]
\renewcommand{\arraystretch}{1}
\begin{center}
\scalebox{1}[1]{
\begin{tabular}{c||c c|c c c c | c c c c | c  }
object & $t$ & $x^1$ & $r$ & $\theta^{1}$ & $\theta^{2}$ & $y$  & $\rho$ & $\varphi^1$ & $\varphi^2$ & $\phi$ & $z$    \\
\hline \hline
$\mrm{M}5_\xi$ & $\times$ & $\times$ & $\times$ & $\times$ & $\times$ & $\times$ & $-$ & $-$ & $-$ & $-$ & $-$ \\
\hline
$\mrm{M}2$& $\times$ & $\times$ & $-$ & $-$ & $-$ & $-$  & $\sim$ & $\sim$ & $\sim$ & $\sim$ & $\times$ \\
$\mrm{M}5$ & $\times$ & $\times$ & $-$ & $-$ & $-$ & $-$ & $\times$ & $\times$ & $\times$ & $\times$ & $\sim$\\
$\mrm{KK}$ & $\times$ & $\times$ &$-$ & $-$ & $-$ & $\text{ISO}$ & $\times$ & $\times$ & $\times$  & $\times$ & $\times$\\
\end{tabular}
}
\end{center}
\caption{$\frac18$ -- BPS brane setup of defect M2 -- M5 branes at a conical singularity intersecting orthogonal M5$_\xi$ branes with KK monopoles. This background describes the supergravity dual of $\ma N=(0,4)$ a surface defect within the 6d $\ma N=(1,0)$ SCFT dual to AdS$_7/\mathbb{Z}_k$ vacua of M-theory.} \label{Table:M2M5LambertKK}
\end{table}
After the inclusion of KK monopoles in the string background \eqref{lambertdefect}, we obtains the following background 
\be
\label{AdS3_S3_f_KK}
\begin{array}{lclc}
ds_{11}^{2} & = &  f^{2/3}\,\left(H^{-2/3}\left(s^2+c^2H\right)^{1/3}L^2\,ds_{\mathrm{AdS}_3}^2+ \,H^{1/3}\left(s^2+c^2H\right)^{-2/3} \kappa^2\,ds_{S^3/\mathbb{Z}_k}^2\right)\,+\\[1mm]
& + &  f^{-1/3}H^{1/3}\,\left(s^2+c^2H\right)^{1/3}\,\left(ds^2_{M^4} + f^{-1} \, dz^2\right) .&  \\[3mm]
\end{array}
\ee
 It is easy to show that if $M_4=T^4$, the aforementioned background turns out to be included in the classification\footnote{The general solution (B.1) of appendix B in \cite{Lozano:2020bxo} includes our backgrounds \eqref{AdS3_S3_f_KK}. In particular by fixing $\mrm{CY}_2=T^4$ and by considering the connection $\ma A$ vanishing, it turns out that our linear function $f(z)$ plays the role of the function $u$ introduced therein. The quantity $(u^\prime)^2$ turns out to be expressed in our notation in terms of the parameter $Q$ given in \eqref{xi2Q_relation}, hence in terms of $\tan \xi$. Finally our function $H$ is the $h_4$ function of \cite{Lozano:2020bxo}.} of $\ma N=(0,4)$ $\mrm{AdS}_3\times S^3/\mathbb{Z}_k\times \mrm{CY}_2\times I$ solutions in M-theory of \cite{Lozano:2020bxo} once we choose $M_4=\mrm{CY}_2=T^4$. In particular the backgrounds \eqref{AdS3_S3_f} provide a clear brane interpretation of those solutions in \cite{Lozano:2020bxo} featured by a warping within $\mrm{AdS}_3$ and $S^3/\mathbb{Z}_k$ that, in that context, it is encoded in a linear function called $u$. We conclude by mentioning that providing a clear interpretation as near-horizon of brane solution of those $\mrm{AdS}_3$ solutions in \cite{Lozano:2020bxo} characterized by $\mrm{CY}_2=K3$, remains an open problem.

\section*{Acknowledgements}

We would like to thank Christopher Couzens, Yolanda Lozano, Niall Macpherson and Carlos Nu$\tilde{\textrm{n}}$ez for very interesting and stimulating discussions.  
The work of GD is supported by the STARS grant named THEsPIAN.
The work of NP is supported by the Principado de Asturias through the grant FC-GRUPIN-IDI/2018/000174.


 \bibliographystyle{utphys}
  \bibliography{references}

\providecommand{\href}[2]{#2}\begingroup\raggedright\begin{thebibliography}{10}

\bibitem{Maldacena:1997re}
J.~M. Maldacena, ``{The Large N limit of superconformal field theories and
  supergravity},'' \href{http://dx.doi.org/10.1023/A:1026654312961}{{\em Int.
  J. Theor. Phys.} {\bfseries 38} (1999) 1113--1133},
  \href{http://arxiv.org/abs/hep-th/9711200}{{\ttfamily arXiv:hep-th/9711200}}.

\bibitem{Witten:1998qj}
E.~Witten, ``{Anti-de Sitter space and holography},''
  \href{http://dx.doi.org/10.4310/ATMP.1998.v2.n2.a2}{{\em Adv. Theor. Math.
  Phys.} {\bfseries 2} (1998) 253--291},
  \href{http://arxiv.org/abs/hep-th/9802150}{{\ttfamily arXiv:hep-th/9802150}}.

\bibitem{Aharony:2008ug}
O.~Aharony, O.~Bergman, D.~L. Jafferis, and J.~Maldacena, ``{N=6 superconformal
  Chern-Simons-matter theories, M2-branes and their gravity duals},''
  \href{http://dx.doi.org/10.1088/1126-6708/2008/10/091}{{\em JHEP} {\bfseries
  10} (2008) 091}, \href{http://arxiv.org/abs/0806.1218}{{\ttfamily
  arXiv:0806.1218 [hep-th]}}.

\bibitem{Cvetic:2000cj}
M.~Cvetic, H.~Lu, C.~Pope, and J.~F. Vazquez-Poritz, ``{AdS in warped
  space-times},'' \href{http://dx.doi.org/10.1103/PhysRevD.62.122003}{{\em
  Phys. Rev. D} {\bfseries 62} (2000) 122003},
  \href{http://arxiv.org/abs/hep-th/0005246}{{\ttfamily arXiv:hep-th/0005246}}.

\bibitem{Strominger:1986uh}
A.~Strominger, ``{Superstrings with Torsion},''
  \href{http://dx.doi.org/10.1016/0550-3213(86)90286-5}{{\em Nucl. Phys. B}
  {\bfseries 274} (1986) 253}.

\bibitem{Gauntlett:2002fz}
J.~P. Gauntlett and S.~Pakis, ``{The Geometry of D = 11 killing spinors},''
  \href{http://dx.doi.org/10.1088/1126-6708/2003/04/039}{{\em JHEP} {\bfseries
  04} (2003) 039}, \href{http://arxiv.org/abs/hep-th/0212008}{{\ttfamily
  arXiv:hep-th/0212008}}.

\bibitem{Grana:2005sn}
M.~Grana, R.~Minasian, M.~Petrini, and A.~Tomasiello, ``{Generalized structures
  of N=1 vacua},'' \href{http://dx.doi.org/10.1088/1126-6708/2005/11/020}{{\em
  JHEP} {\bfseries 11} (2005) 020},
  \href{http://arxiv.org/abs/hep-th/0505212}{{\ttfamily arXiv:hep-th/0505212}}.

\bibitem{Karch:2000gx}
A.~Karch and L.~Randall, ``{Open and closed string interpretation of SUSY CFT's
  on branes with boundaries},''
  \href{http://dx.doi.org/10.1088/1126-6708/2001/06/063}{{\em JHEP} {\bfseries
  06} (2001) 063}, \href{http://arxiv.org/abs/hep-th/0105132}{{\ttfamily
  arXiv:hep-th/0105132}}.

\bibitem{Argurio:2000tg}
R.~Argurio, A.~Giveon, and A.~Shomer, ``{Superstring theory on AdS(3) x G / H
  and boundary N=3 superconformal symmetry},''
  \href{http://dx.doi.org/10.1088/1126-6708/2000/04/010}{{\em JHEP} {\bfseries
  04} (2000) 010}, \href{http://arxiv.org/abs/hep-th/0002104}{{\ttfamily
  arXiv:hep-th/0002104}}.

\bibitem{Kim:2005ez}
N.~Kim, ``{AdS(3) solutions of IIB supergravity from D3-branes},''
  \href{http://dx.doi.org/10.1088/1126-6708/2006/01/094}{{\em JHEP} {\bfseries
  01} (2006) 094}, \href{http://arxiv.org/abs/hep-th/0511029}{{\ttfamily
  arXiv:hep-th/0511029}}.

\bibitem{Gauntlett:2006ns}
J.~P. Gauntlett, N.~Kim, and D.~Waldram, ``{Supersymmetric AdS(3), AdS(2) and
  Bubble Solutions},''
  \href{http://dx.doi.org/10.1088/1126-6708/2007/04/005}{{\em JHEP} {\bfseries
  04} (2007) 005}, \href{http://arxiv.org/abs/hep-th/0612253}{{\ttfamily
  arXiv:hep-th/0612253}}.

\bibitem{Gauntlett:2006af}
J.~P. Gauntlett, O.~A. Mac~Conamhna, T.~Mateos, and D.~Waldram,
  ``{Supersymmetric AdS(3) solutions of type IIB supergravity},''
  \href{http://dx.doi.org/10.1103/PhysRevLett.97.171601}{{\em Phys. Rev. Lett.}
  {\bfseries 97} (2006) 171601},
  \href{http://arxiv.org/abs/hep-th/0606221}{{\ttfamily arXiv:hep-th/0606221}}.

\bibitem{Donos:2008hd}
A.~Donos, J.~P. Gauntlett, and J.~Sparks, ``{AdS(3) x (S**3 x S**3 x S**1)
  Solutions of Type IIB String Theory},''
  \href{http://dx.doi.org/10.1088/0264-9381/26/6/065009}{{\em Class. Quant.
  Grav.} {\bfseries 26} (2009) 065009},
  \href{http://arxiv.org/abs/0810.1379}{{\ttfamily arXiv:0810.1379 [hep-th]}}.

\bibitem{DHoker:2008rje}
E.~D'Hoker, J.~Estes, M.~Gutperle, and D.~Krym, ``{Exact Half-BPS Flux
  Solutions in M-theory II: Global solutions asymptotic to AdS(7) x S**4},''
  \href{http://dx.doi.org/10.1088/1126-6708/2008/12/044}{{\em JHEP} {\bfseries
  12} (2008) 044}, \href{http://arxiv.org/abs/0810.4647}{{\ttfamily
  arXiv:0810.4647 [hep-th]}}.

\bibitem{Couzens:2017way}
C.~Couzens, C.~Lawrie, D.~Martelli, S.~Schafer-Nameki, and J.-M. Wong,
  ``{F-theory and AdS$_{3}$/CFT$_{2}$},''
  \href{http://dx.doi.org/10.1007/JHEP08(2017)043}{{\em JHEP} {\bfseries 08}
  (2017) 043}, \href{http://arxiv.org/abs/1705.04679}{{\ttfamily
  arXiv:1705.04679 [hep-th]}}.

\bibitem{Eberhardt:2017uup}
L.~Eberhardt, ``{Supersymmetric AdS$_{3}$ supergravity backgrounds and
  holography},'' \href{http://dx.doi.org/10.1007/JHEP02(2018)087}{{\em JHEP}
  {\bfseries 02} (2018) 087}, \href{http://arxiv.org/abs/1710.09826}{{\ttfamily
  arXiv:1710.09826 [hep-th]}}.

\bibitem{Dibitetto:2017tve}
G.~Dibitetto and N.~Petri, ``{BPS objects in D = 7 supergravity and their
  M-theory origin},'' \href{http://dx.doi.org/10.1007/JHEP12(2017)041}{{\em
  JHEP} {\bfseries 12} (2017) 041},
  \href{http://arxiv.org/abs/1707.06152}{{\ttfamily arXiv:1707.06152
  [hep-th]}}.

\bibitem{Dibitetto:2018ftj}
G.~Dibitetto, G.~Lo~Monaco, A.~Passias, N.~Petri, and A.~Tomasiello, ``{AdS$_3$
  Solutions with Exceptional Supersymmetry},''
  \href{http://dx.doi.org/10.1002/prop.201800060}{{\em Fortsch. Phys.}
  {\bfseries 66} no.~10, (2018) 1800060},
  \href{http://arxiv.org/abs/1807.06602}{{\ttfamily arXiv:1807.06602
  [hep-th]}}.

\bibitem{Couzens:2019iog}
C.~Couzens, ``{$\mathcal{N}=(0,2)$ AdS$_3$ Solutions of Type IIB and F-theory
  with Generic Fluxes},'' \href{http://arxiv.org/abs/1911.04439}{{\ttfamily
  arXiv:1911.04439 [hep-th]}}.

\bibitem{Lozano:2019emq}
Y.~Lozano, N.~T. Macpherson, C.~Nunez, and A.~Ramirez, ``{AdS$_3$ solutions in
  Massive IIA with small $\mathcal{N}=(4,0)$ supersymmetry},''
  \href{http://dx.doi.org/10.1007/JHEP01(2020)129}{{\em JHEP} {\bfseries 01}
  (2020) 129}, \href{http://arxiv.org/abs/1908.09851}{{\ttfamily
  arXiv:1908.09851 [hep-th]}}.

\bibitem{Lozano:2020bxo}
Y.~Lozano, C.~Nunez, A.~Ramirez, and S.~Speziali, ``{$M$-strings and AdS$_3$
  solutions to M-theory with small $\mathcal{N}=(0,4)$ supersymmetry},''
  \href{http://dx.doi.org/10.1007/JHEP08(2020)118}{{\em JHEP} {\bfseries 08}
  (2020) 118}, \href{http://arxiv.org/abs/2005.06561}{{\ttfamily
  arXiv:2005.06561 [hep-th]}}.

\bibitem{Faedo:2020nol}
F.~Faedo, Y.~Lozano, and N.~Petri, ``{Searching for surface defect CFTs within
  AdS$_3$},'' \href{http://arxiv.org/abs/2007.16167}{{\ttfamily
  arXiv:2007.16167 [hep-th]}}.

\bibitem{Izquierdo:1995ms}
J.~Izquierdo, N.~Lambert, G.~Papadopoulos, and P.~Townsend, ``{Dyonic
  membranes},'' \href{http://dx.doi.org/10.1016/0550-3213(95)00606-0}{{\em
  Nucl. Phys. B} {\bfseries 460} (1996) 560--578},
  \href{http://arxiv.org/abs/hep-th/9508177}{{\ttfamily arXiv:hep-th/9508177}}.

\end{thebibliography}\endgroup
\end{document}